\documentclass[a4paper,11pt]{article}
\pdfoutput=1 

\usepackage{jcappub} 

\usepackage[T1]{fontenc} 

\usepackage{amsmath,amssymb,amsfonts}
\usepackage{graphicx,graphics,epsfig}
\usepackage{enumerate} 
\usepackage{colordvi} 
\usepackage{bm}
\usepackage{epsfig}
\usepackage{float}
\usepackage{verbatim}
\usepackage{subfig}
\usepackage{footnote}

\usepackage[dvipsnames]{xcolor}

\usepackage[colorinlistoftodos]{todonotes}
\usepackage{hyperref}

\newcommand{\be}{\begin{equation}}
\newcommand{\ee}{\end{equation}}
\newcommand{\bea}{\setlength\arraycolsep{2pt} \begin{eqnarray}}
\newcommand{\eea}{\end{eqnarray}}

\def\0{{\sst{(0)}}}
\def\1{{\sst{(1)}}}
\def\2{{\sst{(2)}}}
\def\3{{\sst{(3)}}}
\def\4{{\sst{(4)}}}
\def\5{{\sst{(5)}}}
\def\6{{\sst{(6)}}}
\def\7{{\sst{(7)}}}
\def\8{{\sst{(8)}}}
\def\sst#1{{\scriptscriptstyle #1}}

\title{Baryogenesis through Asymmetric Hawking Radiation from Primordial Black Holes as Dark Matter} 

\author[a,b,c]{Alexis~Boudon,}
\author[a]{Benjamin~Bose,}
\author[d,e,a]{Hyat~Huang,}
\author[a]{Lucas~Lombriser}

\affiliation[a]{D\'epartement de Physique Th\'eorique, Universit\'e de Gen\`eve, 24~quai Ernest Ansermet, 1211~Gen\`eve~4, Switzerland}
\affiliation[b]{D\'epartement de Physique, Universit\'e Claude Bernard Lyon~1, 14~rue Enrico Fermi, 69622 Villeurbanne, France}
\affiliation[c]{Institut de Physique Th\'eorique, CEA Paris-Saclay, Bat.~774~Ormes des Merisiers, 91191 Gif-sur-Yvette, France}
\affiliation[d]{College of Physics and Communication Electronics, Jiangxi Normal University, Nanchang 330022, China}
\affiliation[e]{Department of Physics, Beijing Normal University, Beijing 100875, China}

\emailAdd{alexis.boudon@etu.univ-lyon1.fr}
\emailAdd{benjamin.bose@unige.ch}
\emailAdd{hyat@mail.bnu.edu.cn}
\emailAdd{lucas.lombriser@unige.ch}

\abstract{
We examine the extent to which primordial black holes (PBHs) can constitute the observed dark matter while also giving rise to the measured matter-antimatter asymmetry and account for the observed baryon abundance through asymmetric Hawking radiation generated by a derivative coupling of curvature to the baryon-lepton current. We consider both broad and monochromatic mass spectra for this purpose. For the monochromatic spectrum we find that the correct dark matter and baryon energy densities are recovered for peak masses of the spectrum of $M_{\rm pk} \geq 10^{12}$~kg whereas for the broad case the observed energy densities can be reproduced regardless of peak mass. Adopting some simplifications for the early-time expansion history as a first approximation, we also find that the measured baryon asymmetry can be recovered within an order of magnitude. We argue furthermore that the correct value of the baryon-lepton yield can in principle be retrieved for scenarios where a significant amount of the radiation is produced by PBH decay during or after reheating, as is expected when the decaying PBHs also cause reheating, or when an early matter-dominated phase is considered. We conclude from this first analysis that the model merits further investigation.
}
\begin{document}
\maketitle
\flushbottom

\section{Introduction}

The standard model of Cosmology ($\Lambda$CDM) stipulates a homogeneous and isotropic Universe with vanishing spatial curvature that contains two energy components in addition to the ordinary baryonic matter: cold dark matter (CDM) and the cosmological constant ($\Lambda$).
Despite the observational success of $\Lambda$CDM, there remain significant gaps in our physical understanding of it.
For instance, we have no explanation for the baryon asymmetry, meaning why there is more matter than antimatter in the Universe.
Even more obscure is the nature of the dark matter and the cosmological constant.
The first hypotheses of dark matter date back more than a century, and it has been invoked several times since and in different contexts~\cite{deSwart:2017heh}.
It is only over the past few decades, however, that we have gathered a substantial amount of evidence supporting its existence, ranging from observations on the scales of dwarf galaxies to the extension of the observable Universe from early to late times.
Examples of this evidence include the rotation curves and velocity dispersions in galactic-scale bound systems~\cite{1998gaas.book.....B}, excessive galaxy cluster masses~\cite{Tozzi}, or the baryon acoustic oscillations (BAO)~\cite{glazebrook2005dark, Eisenstein_2005, cole20052df}.
Exquisite measurements of the fractional cosmological energy density in CDM is offered among others by the cosmic microwave background (CMB) radiation observed with \emph{Planck} 2018~\cite{Aghanim:2018eyx}, the Baryonic Oscillation Spectroscopic Survey (BOSS) galaxy clustering measurements~\cite{Ivanov:2019pdj}, weak gravitational lensing observations with the Kilo Degree Survey (KiDS)~\cite{Hildebrandt:2016iqg}, or measurements of cluster abundances~\cite{Mana:2013qba}.

Despite this overwhelming observational evidence, the fundamental nature of dark matter remains unknown.
Many dark matter candidates have been put forward as an explanation for it such as weakly interacting massive particles (WIMPs), axion-like particles, or sterile neutrinos~\cite{roszkowski2017wimp, irastorza2018new, bser2019status}.
But none of these have succeeded in convincingly making their case so far~\cite{lovell2020cosmological, RevModPhys.90.045002}.
A hint for the nature of dark matter may perhaps also be taken from the interesting fact that the baryonic fractional energy density measured by the CMB and Big Bang nucleosynthesis (BBN)~\cite{Inomata:2018htm} happens to lie within the same order of magnitude as that of CDM.
This may suggest some shared generation process that could ultimately also lead to a simultaneous explanation for the baryon asymmetry.

Perhaps the least exotic dark matter candidate, at least for the physical entity involved, that still remains viable is the scenario of primordial black holes (PBHs).
Several physical mechanisms have been proposed to generate PBHs.
For example, they may be produced by large primordial density fluctuations~\cite{Green:2014faa,Crawford:1982yz,DeLuca:2020ioi}, cosmic string loops~\cite{Hawking:1987bn}, or bubble collisions~\cite{Hawking:1982ga}.
This theoretical motivation has naturally also led to their consideration as CDM candidates (see Ref.~\citep{Sasaki:2018dmp} for a review).
The scenario has recently drawn resurged attention due to the direct detection of gravitational waves from black hole binary systems~\citep{Abbott:2016blz,Sasaki:2018dmp,Bird:2016dcv,Clesse:2016vqa,Garcia-Bellido:2017fdg}.
Much of the current research on the PBH-CDM model focuses on determining the PBH abundance in a given mass range, phrased in terms of the \emph{mass spectrum}.
Another important factor to consider is that not all PBHs could have survived until today due to evaporation.
Because of Hawking radiation, PBHs with masses lower than about $4\times 10^{11}$~kg would have evaporated by today.

The fact that PBHs undergo a decay prompts an interesting question: \emph{Could PBHs be subject to an asymmetric Hawking radiation that gives rise to the baryon asymmetry yet allows abundant PBHs to remain to constitute the dark matter?}
Such a scenario may also naturally provide a link between the coincident baryon and dark matter abundances. Hawking radiation of black holes has indeed been invoked as a generator of baryon number asymmetry~\citep{Hawking:1974rv}.
An asymmetric decay can, for example, arise from the presence of a non-vanishing chemical potential, which can for instance be the result of a derivative coupling of curvature to the baryon-lepton current~\citep{Hook:2014mla, Hamada_2017} and lead to the radiation of more particles than anti-particles.
More generally, solving baryogenesis through gravitational effects is known as gravitational baryogenesis and can be realised through different mechanisms~\citep{Davoudiasl:2004gf}.
In this paper, we will investigate the feasibility of PBHs as CDM candidates that partially decay to produce the observed baryon asymmetry.
For this purpose, we will adopt the gravitational baryogenesis model of Ref.~\citep{Hook:2014mla} and generalise it to extended PBH mass spectra. 
Specifically, we consider both monochromatic~\cite{Carr:2016drx} and polychromatic~\cite{DeLuca:2020ioi} mass spectra.
For those we compare the predicted fractional cosmological energy density in PBHs with that observed for CDM. Then, using the asymmetric Hawking radiation mechanism of Ref.~\cite{Hook:2014mla} we estimate the number density of baryons in the Universe and the baryon abundance, which we compare against observations. Finally, we also estimate the baryon-lepton yield, which quantifies the amount of matter-antimatter asymmetry in the Universe and compare that to observations.

The paper is laid out as follows: in Sec.~\ref{sec:model} we give a very brief review for the case of PBHs as dark matter and describe the PBH mass spectra and the model for asymmetric Hawking radiation we will use. Sec.~\ref{sec:predictions} derives the model predictions: the PBH and baryon fractional energy densities and the baryon-lepton yield. In Sec.~\ref{sec:observations} we compare our predictions against observations of the CMB and the large-scale structure as well as the observed matter-antimatter asymmetry. We conclude and summarise in Sec.~\ref{sec:summary}.
Finally, we provide some additional details and a discussion of assumptions and extensions of our computations in the appendix.
All quantities in this article are expressed in natural units unless explicitly stated.


\section{Baryogenesis from the decay of Primordial Black Holes as Dark Matter} \label{sec:model}
We will first provide a brief review of the theoretical models we will adopt in our analysis before presenting our predictions and the comparison to observations in Secs.~\ref{sec:predictions} and \ref{sec:observations}.
In particular, we will be considering PBHs as a CDM candidate with a given mass spectrum (Secs.~\ref{sec:pbhascdm} and \ref{sec:spectrum}).
These PBHs will then radiate asymmetric Hawking radiation to produce the observed baryon fraction and asymmetry (Sec.~\ref{sec:baryogenesis}).


\subsection{Primordial Black Holes as Dark Matter}\label{sec:pbhascdm}

The latest measurements of the \emph{Planck} collaboration constrain the present day CDM energy density fraction to $\Omega_{\rm CDM,0} = 0.267$~\cite{Akrami:2018vks,Aghanim:2018eyx}, assuming a flat $\Lambda$CDM scenario.
Attempts to address this significant contribution to the cosmic energy budget with PBHs can be traced back to the works of  Zeldovich and Novikov in 1967~\cite{novi} and Hawking in 1971~\cite{Hawking:1971ei}, who first proposed their existence. The ability of PBHs to describe the observed CDM depends on their abundance and mass range. 

The mass of PBHs is related to their formation time. Compared with the black holes that are formed from the collapse of stars, PBHs have a broader mass range. This is due to the fact that the formation of PBHs does not need to satisfy the Tolman–Oppenheimer–Volkoff limit. To form, PBHs need a density fluctuation of $\delta\rho\sim0.1$~\cite{Harada:2013epa}. Thus we usually expect that PBHs were produced between the era of inflation to just before the end of reheating. Given reheating must occur before BBN, this implies the possibility of their formation from $10^{-34}$~s to at least $10$~s~\cite{Coc:2017pxv} after the Big Bang. Consequently, PBHs could form in a mass range between the Planck mass, $M_p = 2.18 \times 10^{-8}$~kg, to a maximum mass of $M_{\rm max} \sim 10^{36}$~kg~\cite{Carr:2016drx,Carr:2020gox}.

An upper bound of PBH-CDM masses of $10^{18}$~kg was suggested in Ref.~\citep{Blais:2002nd} from limitations of cosmic gamma ray bursts.
But based on the study of cosmic entropy and black hole entropy, Frampton~\cite{Frampton:2010sw} argued that if all CDM is composed of PBHs, the mass of PBHs can be in the range of $10^{22}$--$10^{35}$~kg.
We direct the reader to Ref.~\citep{Carr:2020gox} for a review on the observational constraints on PBHs as a CDM candidate.

One important assumption in these constraints is that a single population of PBHs of a given mass, or narrow mass range, makes up all the CDM. These scenarios are characterised by a so-called \emph{monochromatic} mass spectrum. Such spectra must peak in narrow mass gaps, e.g., $10^{13}$--$10^{14}$~kg~\cite{Carr:2016drx}, for them to account for all of the CDM fraction. Recently, mass spectra covering a very broad range of masses have been studied in Ref.~\cite{DeLuca:2020ioi}. Such a \emph{polychromatic} spectrum may be able to more easily evade some of the observational constraints~\cite{Carr:2016drx}. We shall discuss the mass spectrum of PBHs next.


\subsection{The Primordial Black Hole spectrum} \label{sec:spectrum}

The PBH mass spectrum can be defined as 
\begin{equation}
    \psi(M) \equiv \frac{1}{\bar{\rho}_{\rm PBH}} \frac{d \rho(M)}{dM} \, ,
    \label{eq:massspecdef}
\end{equation}
where $\bar{\rho}_{\rm PBH}$ is the total matter density of PBHs and $d\rho$ is their matter density in the mass range $(M, M+dM)$.
Black holes that are produced at a \emph{given epoch} in the early Universe are associated with density fluctuations with a delta-function power spectrum~\cite{Yokoyama:1998xd,Niemeyer:1999ak,Musco:2012au,Carr:2016hva,Carr:2017jsz}. This produces a very sharp mass spectrum. Such \emph{monochromatic} spectra can take the form 
\begin{equation}
    \psi_{\rm mono}(M,t) = A_\psi(t) M^{2.85}(t) \exp [ - (M(t)/M_{\rm pk})^{2.85}] \, , \label{eq:mono}
\end{equation}
where $M_{\rm pk}$ corresponds to a fraction of the horizon mass at the collapse epoch, which roughly evolves as~\cite{Carr:2016drx,Carr:2020gox}
\begin{equation}
    M_{\rm H} \sim 10^{12}\left(\frac{t}{10^{-23}~\rm{s}}\right)~\rm{kg} \, ,
    \label{eq:mf}
\end{equation}
where $t$ is the time coordinate.
$A_\psi(t)$ is a normalisation constant that ensures the proper definition of $\psi$ as a probability density,
\begin{equation}
    A_\psi(t) = \left( \int_{\rm M_p}^{M_{\rm max}} M^{2.85} \exp [ - (M/M_{\rm pk})^{2.85})] dM \right)^{-1} \, ,
\end{equation}
and has units of ${\rm [kg]}^{-3.85}$, where we set $M_{\rm max} = 10^{36}$~kg.
In practice, for the monochromatic spectrum the maximum mass just needs to be chosen sufficiently higher than the peak because of the sharp fall-off.
With Eq.~\eqref{eq:mono} PBHs are being produced at all masses below the horizon mass but for masses above that, the probability of formation is negligible.

The horizon mass is of course a function of time, and so as the Universe expands, we can produce black holes corresponding to larger and larger masses. One can model the corresponding mass spectrum by considering Eq.~\eqref{eq:mono} for the relevant epochs. The resulting superposition of these functions gives a much broader, \emph{polychromatic}, mass spectrum such as that described in Ref.~\cite{DeLuca:2020ioi},
\begin{equation}
    \psi_{\rm poly, high}(M) = \frac{1}{f_{\rm PBH}} \sqrt{\frac{M_{eq}}{M^3}} \beta(M) \, , \label{eq:polyhm}
\end{equation}
where $M_{eq} = 2.8 \times 10^{17}M_\odot$ is the horizon mass at the time of radiation-matter equality ($t\sim 10^{12}$~s), $f_{\rm PBH}$ is the fraction of CDM that the PBHs constitute and $\beta(M)$ is the abundance of PBHs of mass M, which is constant in the high mass regime \cite{DeLuca:2020ioi}. We are interested in $f_{\rm PBH} \approx 1 $.  In practice, to get a function, $\psi_{\rm poly}$, for the full spectrum of masses, one can sum Eq.~\eqref{eq:mono} over many masses starting with the mass corresponding to when the shortest wavelength mode re-enters the horizon, $M_s$, and ending with the longest, $M_l$. 

\begin{figure}[t]
\centering
  \includegraphics[width=0.85\textwidth]{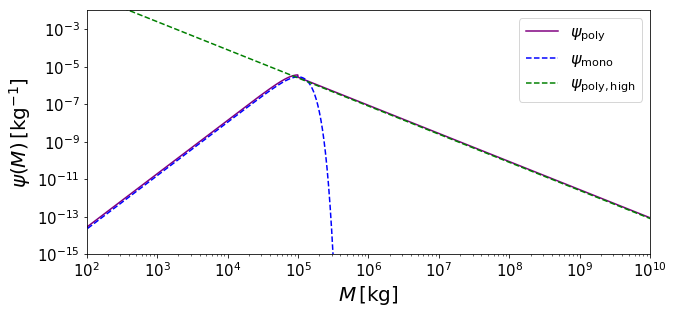}
  \caption[]{Polychromatic PBH mass spectrum (purple solid) and a monochromatic spectrum peaked at $M_{\rm pk} = 10^{5}$~kg (blue dashed). The polychromatic spectrum is the envelope of many such monochromatic spectra which are produced at different collapse times. The green dashed line illustrates the behaviour of the polychromatic spectrum at the high-mass end given by Eq.~\eqref{eq:polyhm}.} 
\label{fig:massfunc}
\end{figure}

In this work, we approximate the full broad spectrum by interpolating between Eqs.~\eqref{eq:mono} and \eqref{eq:polyhm} by  tuning the value of $\beta$. Because of this, we take $M_{\rm pk}$ to be a free parameter, corresponding to a choice of $M_s$. Note that in the full calculation, $M_s$ will be close to $M_{\rm pk}$. We then normalise the distribution so that
\begin{equation}
    \int_{M_p}^{M_{\rm l}} \psi_{\rm poly} (M_i,t) dM_i = 1 \,,
\end{equation}
where $M_i$ refers to the initial mass of a PBH at the time it is created. We have taken $M_l \approx {\rm Min}[M_H(t),10^{36}~\rm{kg}]$ (see Eq.~\ref{eq:mf}), which assumes that PBHs can form up to just before BBN~\cite{Coc:2017pxv}. We find that our results are largely insensitive to the choice of maximum mass and direct the reader to App.~\ref{app:assext} for a short discussion.
Figure~\ref{fig:massfunc} shows the monochromatic spectrum given by Eq.~\eqref{eq:mf} with $M_{\rm pk} = 10^5$~kg, the high-mass end spectrum given in Eq.~\eqref{eq:polyhm}, and a final spectrum comprised of PBH populations that form at different collapse epochs. 

\begin{figure}[t]
\centering
  \includegraphics[width=0.85\textwidth]{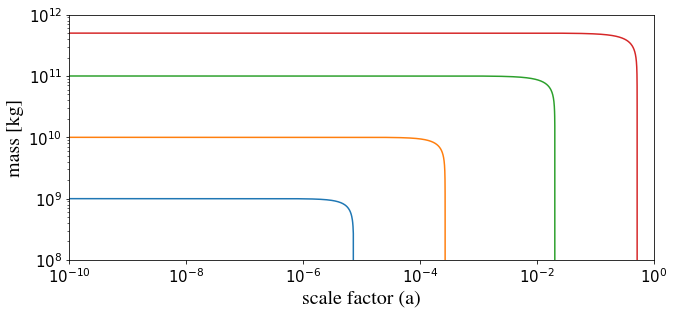}
  \caption[]{Evolution of the PBH mass parametrised by the scale factor $a$ (normalised to unity today). Different initial masses evaporate at different points in the evolution of the Universe.
  } 
\label{fig:bhdecay}
\end{figure}

Finally, each PBH will also decay once created through the emission of Hawking radiation. The decay for Schwarzschild black holes is given by
\cite{Hook:2014mla} 
\begin{equation}
    M(t) = (M_i^3 - 3KM_p^4 t)^{1/3} \Theta\left[t_f(M_i,t_i)-t\right] \,,
    \label{eq:massloss}
\end{equation}
where $K= g_\star / (30720 \pi) $ with $g_\star = \sum_{\rm bosons}g_i + \frac{7}{8} \sum_{\rm fermions} g_i = 106.75$ and $g_i$ is the number of degrees of freedom for the respective particles.
These quantities are further discussed in App.~\ref{sec:dof}. Furthermore, $t_i$ denotes the intial time, $\Theta(t_f-t)$ is the Heaviside step function, and $t_f(M_i,t_i) = (M_i/M_p)^3/(3K M_p)+t_i$ denotes the time when the PBH has completely decayed. Figure~\ref{fig:bhdecay} shows that black holes that are produced with masses less than $M\approx 4\times 10^{11}$~kg will have completely decayed by today~\cite{Carr:2009jm}.

\subsection{Baryogenesis from asymmetric Hawking radiation} \label{sec:baryogenesis} 

From the measurement of the CMB radiation, we infer an asymmetry between particles and anti-particles in our Universe of $n_B/s \approx 8.7 \times 10^{-11}$~\cite{Planck2016}, where $n_B$ is the number density of baryons and $s$ is the entropy density. Sakharov~\cite{Sakharov:1967dj} proposed three conditions for a process to produce this asymmetry:
\begin{enumerate}
    \item 
    The baryon number density, $n_B$, must not be conserved.  
    \item
    C- and CP-symmetry must be violated in order for the process to produce more baryons than anti-baryons.
    \item
    The process occurs out of thermal equilibrium. 
\end{enumerate}

In Ref.~\cite{Hook:2014mla} a dynamically generated chemical potential was proposed as an explanation for the baryon asymmetry that arises from the effect of the expansion of the Universe on a CP-violating coupling of the form
\begin{equation}
    \mathcal{S} \supset \int dx^4 \sqrt{-g} \lambda \frac{\partial_\mu R} {M_p^2} J^{\mu}_{\rm B-L} \,,
    \label{eq:lagrangian}
\end{equation}
where $R$ is the Ricci scalar, $\lambda$ is a dimensionless coupling constant, and $J^{\mu}_{\rm B-L}$ is the baryon-lepton current.
Note that it is not necessary for this to be the baryon current since any current that leads to a net B-L charge is sufficient. This action can give rise to asymmetric Hawking radiation that produces a baryon number. Specifically, it allows for the production of a non-vanishing baryon-lepton charge that sources the asymmetry. We will discuss the mechanisms in more detail in Sec.~\ref{sec:blcharge}.


\section{Model predictions}\label{sec:predictions}

We shall now derive the predictions of the scenario proposed in Sec.~\ref{sec:model} for three key quantities, namely the total cosmological fractional energy density in PBHs $\Omega_{\rm PBH}$ (Sec.~\ref{sec:enden}), the total cosmological fractional energy density in baryonic matter $\Omega_{\rm b}$ (Sec.~\ref{sec:endenbar}), and the baryon-lepton yield (Sec.~\ref{sec:barlepyield}) $Y_{\rm B-L}$, which are all functions of time.


\subsection{Fractional energy density in Primordial Black Holes}\label{sec:enden}

The first quantity which we wish to calculate is the PBH fractional energy density $\Omega_{\rm PBH}$ given an initial distribution of black holes characterised by the mass spectrum $\psi$. We begin by noting that the initial matter density is given by 
\begin{equation}
    \rho_{\rm PBH, i} = \bar{n}_{\rm PBH, i} \langle M_i \rangle \, , \label{eq:intialdens}
\end{equation}
where $\bar{n}_{\rm PBH, i}$ is the total initial number density of PBHs and the average mass (at any given time) is obtained from
\begin{equation}
 \langle M(t) \rangle = \int_{M_p}^{M_{\rm fin}} M(t) \psi(M_i) dM_i \,, \label{eq:avgmass}
\end{equation}
where $M(t)$ is specified by Eq.~\eqref{eq:massloss}.
We integrate from $M_{\rm p}$ to $M_{\rm fin}= M(t | M_i = 10^{36}~\rm{kg})$ for the monochromatic spectrum (see Eq.~\ref{eq:massloss}). For the polychromatic spectrum we must be a bit more careful since PBHs are produced from the end of inflation well into radiation domination. We take $M_{\rm fin} = {\rm Min}[M_H(t),M(t | M_i = 10^{36}~\rm{kg})]$ for the polychromatic spectrum, taking into account this time dependence. Note that a full treatment of $\psi_{\rm poly}$ would require the following definition for the average mass,
\begin{equation}
    \langle M(t) \rangle  = \int dM M \psi (M, M_s(t),M_l(t)) \, , 
\end{equation}
where $M_s$ and $M_l$ evolve with time too, $M_s$ being parametrised by $M_{\rm pk}$.

For both the monochromatic and polychromatic cases, we assume that all PBHs are produced before some initial time after which the shape of the mass spectrum remains fixed. For the monochromatic spectrum this initial time shall be specified by an initial scale factor $a_i$. For the polychromatic case, we should take this to be some time in the radiation dominated Universe. For simplicity, we choose $t_i = 10$~s which is just before BBN. We direct the reader to App.~\ref{app:assext} for more discussion on these assumptions. 

The evolution of the matter density is then dictated by the expansion of the Universe as well as the evolution of the PBH average mass as they decay. We can assume that the total number of PBHs is a conserved quantity, $\bar{N}_i$, although at late times some of them will have a final mass of $M_f = 0$ due to Hawking radiation.
The Heaviside function in Eq.~\eqref{eq:massloss} takes care of giving these evaporated PBHs a null weighting in the integral.
Note that in App.~\ref{app:assext} we also consider the case when we have remnants, $M_f=M_p$, instead.
With these specifications, we can rewrite Eq.~\eqref{eq:intialdens} at a given time as 
\begin{equation}
    \rho_{\rm PBH}(a) = \frac{\bar{N}_{i}}{V(a)} \langle M (a) \rangle  = \frac{\bar{n}_{\rm 0}}{a^3} \langle M (a) \rangle\, , \label{eq:genden}
\end{equation}
where we have used the scale factor $a$ to parametrise the evolution and $\bar{n}_{\rm 0}=\bar{n}_{\rm PBH, i} a_i^3$ denotes the number density of PBHs today.
Note that we normalise the scale factor today, $a_0=1$, and so $V(a)= V_0 a^3$, where $V_0$ is the current volume of the Universe.

We can now calculate the PBH fraction of the cosmic energy density as
\begin{equation}
 \Omega_{\rm PBH}(t)  =  \frac{\bar{n}_0}{\rho_c(t)}\left(\frac{1}{a}\right)^3 \langle M(t) \rangle \, , \label{eq:BHdensity}
\end{equation}
where $\rho_c=  3H^2/(8\pi G)$ is the critical density. We have assumed here that the PBH density decays as the volume of the Universe as is typical for non-relativistic matter species.
This means we neglect the small impact of the contribution of newly formed black holes or the effect of mergers.
We again direct the reader to App.~\ref{app:assext} for further details.

One important point here is the choice of background evolution $H(a)$ used in the calculation of the critical density $\rho_c(a)$. Observational data does not allow us to deviate strongly from a $\Lambda$CDM background expansion and so we use $H(a)$ with the observed energy density fractions as measured by the \emph{Planck} satellite~\cite{Akrami:2018vks,Aghanim:2018eyx}. This is discussed further in Sec.~\ref{sec:blcharge}.


\subsection{The baryon-lepton yield} \label{sec:barlepyield}

We shall now predict the baryon-lepton yield defined as
\begin{equation}
    Y^{\rm B-L} \equiv \frac{\bar{n}_{\rm B-L}}{s} \, , \label{yield}
\end{equation}
where $n_{\rm B-L}$ is the baryon-lepton number density and $s$ is the entropy density of the Universe. This can be related to the PBH yield and consequently to the total number density of PBHs and their masses.

In Ref.~\cite{Hook:2014mla}, a single population of PBHs of masses $M$ was considered, which can be phrased as a delta-funtion mass spectrum, $\psi(M, M_{\rm pk}) = \delta_D(M-M_{\rm pk})$ with $M=M_{\rm pk}$. When placed in an expanding Universe with a chemical potential, these black holes generate $n_{\rm B-L}$ from
\begin{equation}
    \frac{d Y_{\rm B-L}}{dt} \equiv \frac{d (\bar{n}_{\rm B-L}/s)}{dt} = \frac{d Q}{dt} Y_{\rm PBH} = \frac{dQ}{dt} \frac{n_{\rm PBH}}{s} \,,
\end{equation}
where $n_{\rm PBH}$ is the number density of PBHs of masses $M$ and $Q_{\rm B-L}(M,t)$ is the total charge asymmetry generated from the decay of a PBH of mass $M$.
Both quantities are functions of time.
As we shall see, the quantity $n_{\rm PBH}/s$ is approximately constant for large peak masses and so one can then express the number density of B-L as 
\begin{equation}
    n_{\rm B-L}(t) \approx Q_{\rm B-L}(M,t)  n_{\rm PBH}(t) \, . \label{eq:blasym}
\end{equation}

We will generalise Eq.~\eqref{eq:blasym} from a single-mass population to become applicable for the PBH mass spectra $\psi_{\rm mono}$ and $\psi_{\rm poly}$ described in Sec.~\ref{sec:spectrum}.


\subsubsection{Baryon-lepton charge} \label{sec:blcharge}

To generate asymmetric Hawking radiation, there must be a chemical potential $\mu_i(t)$, which we will take as being generated dynamically through the effect of cosmic expansion on the coupling in Eq.~\eqref{eq:lagrangian} as in Ref.~\cite{Hook:2014mla}.
Note that the majority of the black hole radiation will be in the form of symmetric particle emission, which will largely constitute radiation emissions through pair annihilation, in addition to direct photon emissions. The rate of production of baryon-lepton charge is given by \cite{Hook:2014mla}
\begin{equation}
    \frac{dQ_{\rm B-L}(M,t)}{dt} = 4 \pi r_{\rm +}^2 \sum_{i} q_i g_i \frac{\mu_i(t) T_{\rm H}^2}{24} = \sum_{i} q_i g_i \frac{\mu_i(t)}{96\pi} \,, \label{eq:blqr}  
\end{equation}
where $r_{\rm +}$ is the Schwarzschild radius of the black hole and we have used the Hawking temperature relation $T_{\rm H}(M) =  M_p^2/(8 \pi M)$ in the second equality. Furthermore, $\sum_i g_i q_{i}^2 = 13$ where  $q_i$ and $g_i$ are the B-L charge and degrees of freedom of the emitted particles, which do not include anti-particles (see App.~\ref{sec:details}). Note that $Q_{\rm B-L}$ has no units. We also note that a typo appearing in Ref.~\cite{Hook:2014mla} introduces a `-' sign in Eq.~\eqref{eq:blqr}.

Integrating Eq.~\eqref{eq:blqr} gives the total charge generated from PBH decay up to some time~$t_*$,
\begin{equation}
Q_{\rm B-L}(M_i,t_*) = \int_{t_i}^{t_*} dt \sum_k \frac{q_k g_k}{96\pi} \mu_k(t) \,,  \label{eq:qblint}
\end{equation}
where we assumed a vanishing charge at initial time. Furthermore, assuming that the chemical potential is slowly varying with time with respect to the black hole decay, one finds
\begin{equation}
    Q_{\rm B-L}(M_i,t_*) = \sum_k \frac{q_k^2 g_k}{96\pi} \mu(t_f) [t_f-t_i] \,, \label{eq:qblmuslow} 
\end{equation}
were we have re-defined the chemical potential $\mu \equiv \mu_k/q_k$.
The final time $t_f$ implicitly depends on the PBH mass since if it has decayed by this time, the charge contribution will be zero at all preceding times. We thus have  
\begin{equation}
    t_f = {\rm Min} [t_*, t_{\rm dec}(M_i)] \, , \label{eq:finaltime}
\end{equation}
where the decay time is given by  $t_{\rm dec} = \frac{M_{\rm i}^3}{3KM_p^4}$ (see Eq.~\ref{eq:massloss}).
The chemical potential $\mu \equiv \mu_k/q_k$ is given by (see Eq.~\ref{eq:lagrangian})~\cite{Hook:2014mla}
\begin{equation}
 \mu = \frac{\lambda }{M_p^2} \dot{R} = \frac{\lambda }{M_p^2} 3H \left[ 4 (H^2)' + (H^2)'' \right]
  = \frac{9 \lambda }{M_p^2} H^3 \left[ (1+w)(1-3w) + w' \right] \,, \label{eq:chempot}
\end{equation}
where we have used that the Ricci scalar for a spatially homogeneous and isotropic metric is given by $R = 6H (2H + H')$ with primes indicating derivatives with respect to $\ln a$ and its cosmic time derivative is $\dot{R} = H R'$.
Furthermore, $(H^2)' = -3[1+w(t)]H^2$ such that $R = 3H^2[(1-3w(t)]$, where $w$ denotes the equation of state. Note that $w'$ is not present in Eq.~(17) of Ref.~\cite{Hook:2014mla}, where instead it was assumed that there is one dominant component such that $w(t) \approx const$. We remind the reader that $\lambda$ is a free coupling constant entering in the Lagrangian of Eq.~\eqref{eq:lagrangian}. We also note that our results will depend on the specific form of this chemical potential and its resulting evolution. This is discussed further in App.~\ref{app:assext}.

We immediately see from Eq.~\eqref{eq:chempot} that during inflation, when the inflaton is dominating the energy density ($w=-1$), then no asymmetric Hawking radiation is possible. By this consideration we need not consider baryogenesis through black hole decay before the end of inflation. For the monochromatic case, we will assume that the asymmetry begins to be produced at $t_{\rm RH}$, the time of reheating, specified in terms of the reheating temperature $T_{\rm RH}$ as
\begin{equation}
    t_{\rm RH} = \sqrt{\frac{5}{\pi^3 g_\star}} \frac{M_p}{T_{\rm RH}^2} \,, \label{eq:trh}
\end{equation}
where we will take $T_{\rm RH}$ as a free parameter in this work.
This amounts to $t_i = t_{\rm RH}$ in Eq.~\eqref{eq:qblint}. For the polychromatic case, the PBHs continue to form into radiation domination. We take $t_i = 10$~s, which is before BBN. For the polychromatic spectrum, this assumes no evaporation of PBHs before the final mass $M_l$ is created. We discuss this more in App.~\ref{app:assext}.

Finally, solving Eq.~\eqref{eq:qblint} requires the specification of the Hubble evolution $H$, which in turn depends on the baryon fraction evolution that in turn depends on $n_{\rm B-L}$ in this scenario, where baryogenesis occurs through the asymmetric Hawking radiation mechanism. The evolution of this fraction will not exactly scale as $\sim a^{-3}$ since we have an additional time dependence from $Q_{\rm B-L}(M_i,t)$. Since we aim at a model that closely approximates a $\Lambda$CDM background history from some time before last scattering until today, given CMB and late-time constraints, we simply adopt
\begin{equation}
    H(a) = H_0 \sqrt{ \frac{\Omega_{m,0}}{a^3} + \frac{\Omega_{\gamma,0}}{a^4} + \Omega_\Lambda} \,, \label{eq:lcdmback}
\end{equation}
which will be retrospectively satisfied by the choice of parameters that closely reproduce this $\Lambda$CDM cosmology.
For the energy density parameters and Hubble constant $H_0$ we adopt the \emph{Planck} 2018 best-fit values (see Table.~\ref{tab:observations}). For $\Omega_{\gamma,0}$ we take the total radiation fraction including massless neutrinos, and so we get $\Omega_{\gamma,0} \approx 9\times 10^{-5}$. Then, the full evolution of $w(a)$ is obtained from
\begin{equation}
    w(a) = -\frac{2 H'}{3 H} - 1 \,.  \label{eq:eos}
\end{equation}
Substituting Eqs.~\eqref{eq:lcdmback} and \eqref{eq:eos} into Eq.~\eqref{eq:chempot} and then integrating Eq.~\eqref{eq:qblint}, we find the following time dependence for $Q(M_i,a)$ expressed in terms of the scale factor,
\begin{equation}
    Q(a_*) \propto \frac{a_f^3 - a_{\rm i}^3}{a_f^3 a_{\rm i}^3} \, , \label{eq:qtimdep} 
\end{equation}
which is approximately $1/a_{\rm i}^3$ for times much later than the initial time ($a_f \gg a_{\rm i}$).
For this limit we also require large enough peak masses so that $t_f = t_*$ in Eq.~\eqref{eq:finaltime} on average. For the monochromatic case, we have $a_{\rm i} = a_{\rm RH}$, which leads to a reheating temperature dependence of the B-L number density to be discussed next. The polychromatic spectrum takes $a_i(t_i=10{\rm \, s})$.


\subsubsection{Baryon-lepton number density} \label{sec:BLnumden}

Consider the infinitesimal contribution to the baryon-lepton number density asymmetry at some time $t_*$ given by 
\begin{equation}
 dn_{\rm B-L}(M_i,t_*) = Q_{\rm B-L}(M_i,t_*) dn_{\rm PBH} \,, \label{eq:BLnumberdensity1}
\end{equation}
where $dn_{\rm PBH}$ is the infinitesimal number density contribution of black holes at $t_*$.
To see this consider a bin of PBH masses between $M_k-\Delta M$ and $M_k$.
The total number of PBHs in all mass bins from $M_0=0$ to $M_k$ is $N(M_k) = \sum_{j=0}^{j=k} \Delta N_j$, where $\Delta N_j = N(M_j) - N(M_j-\Delta M)$ is the number of PBHs in the bin $j$.
The total B-L charge from all mass bins from $M_0=0$ to $M_k$ is $\hat{Q}_{\rm B-L}(M_k) = \sum_{j=0}^{j=k} Q_{\rm B-L}(M_j) \Delta N_j$.
We can also define $\hat{Q}_{\rm B-L}(M_k) = \sum_{j=0}^{j=k} \Delta \hat{Q}_{\rm B-L,j}$.
Hence, $\Delta \hat{Q}_{\rm B-L,j} = Q_{\rm B-L}(M_j) \Delta N_j$.
Dividing by $\Delta M$, and taking the limit of $\Delta M \rightarrow 0$ gives $\frac{d\hat{Q}_{\rm B-L}}{dM}(M_j) = Q_{\rm B-L}(M_j) \frac{dN}{dM}(M_j)$.
Dividing by volume and rewriting $d\hat{Q}_{\rm B-L}$ gives Eq.~\eqref{eq:BLnumberdensity1}.

Integrating Eq.~\eqref{eq:BLnumberdensity1} yields
\begin{align}
 \bar{n}_{\rm B-L}(a_*) & \equiv \int dn_{\rm B-L}(a_*)
 \nonumber \\ 
 & = \int_{\mathbb{R}} Q_{\rm B-L}(M_i,a_*) \frac{dN}{V_i} \frac{V_i}{V_*}
 \nonumber \\ 
 & = \frac{V_i}{V_*} \int_{\mathbb{R}}  Q_{\rm B-L}(M_i,a_*) \frac{dn_i}{dM_i} dM_i \nonumber  \\ 
 & = \bar{\rho}_{PBH,i} \left(\frac{a_i}{a_*}\right)^3 \int_{\mathbb{R}} Q_{\rm B-L}(M_i,a_*) \frac{\psi(M_i)}{M_i}  dM_i  \nonumber \\
  & = \bar{n}_0 \langle M_i \rangle \left(\frac{1}{a_*}\right)^3 \int_{\mathbb{R}} Q_{\rm B-L}(M_i,a_*) \frac{\psi(M_i)}{M_i}  dM_i \, , \label{totalnbl}
\end{align}
where $V_*$ is the volume of the Universe at a scale factor $a_*$ and $V_i$ is the volume at the time of PBH formation. Note that if $\phi(M_i) = \delta_D(M-M_i)$, we have $\bar{n}_{\rm B-L}(t_*) = n_{\rm B-L}$ as given by Eq.~\eqref{eq:blasym}, which reduces to the result of Ref.~\cite{Hook:2014mla} (see Eq.~31 of this reference). 

Furthermore, note that in Eq.~\eqref{totalnbl} we integrate between $M_p$ and $M_{\rm fin}$, where $M_{\rm fin}$ was defined in Sec.~\ref{sec:enden}. Moreover, the initial average mass $\langle M_i \rangle$ is defined at the end of reheating for the monochromatic spectrum but at $t=10$~s for the polychromatic spectrum.


\subsubsection{Entropy density and reheating temperature} 

Using Eqs.~\eqref{totalnbl} and \eqref{yield} we can now proceed to determine the yield.
The entropy density is given by \cite{Hook:2014mla}
\begin{equation}
 s(t_*) = \frac{2\pi^2}{45} g_\star[T(t_*)] T(t_*)^3 \, , \label{eq:entropydens}
\end{equation}
where $g_\star$ is the total particle degrees of freedom, which is a function of temperature. However in this work we take $g_\star$ to be constant and the initial time to be $t_{\rm RH}$, thus $g_\star= 106.75$ (see App.~\ref{sec:dof}). The evolution of the entropy density is then specified by 
\begin{equation}
     s(a) = \frac{2\pi^2}{45} g_\star T(t_{\rm RH})^3 \times \left(\frac{a_{\rm RH}}{a}\right)^3 \, \label{eq:entropydens} \, ,
\end{equation}
where $a_{\rm RH}$ is the scale factor at the time of reheating $t_{\rm RH}$.
We can use Eq.~\eqref{eq:qblint} in Eq.~\eqref{totalnbl} in combination with Eq.~\eqref{eq:entropydens} to compute the yield in Eq.~\eqref{yield}.

\subsection{Fractional energy density in baryons} \label{sec:endenbar}

Finally, to determine the baryonic energy density $\rho_{\rm baryons}$ in the Universe produced through the asymmetric Hawking radiation we use Eq.~\eqref{totalnbl}.
Since the relative amount of antimatter in the Universe is almost zero, we can take $n_{\rm B-L}$ to represent the total number density of baryonic matter.
Then, using the constraints discussed in App.~\ref{app:yieldobs}, we can assume that there is almost the same number of baryons and leptons so that $B \approx -L$. Making the rough assumption that the baryons are all protons and the leptons are all electrons and since the electron mass is negligible compared to that of the proton, the total energy density for baryonic matter is approximately
\begin{equation}
    \rho_{\rm baryons}(t) = \frac{n_{\rm B-L}(t)}{2}(M_{\rm proton} + M_{\rm  electron}) \approx \frac{n_{\rm B-L}(t)}{2}M_{\rm proton} \label{eq:bardens} \, ,
\end{equation} where $M_{\rm electron} = 511$~keV and $M_{\rm proton} = 938$~MeV.
Note that we neglected the number of neutrinos that are generated by PBH evaporation, which may actually be significant, but this is consistent with the primary $Y_{\rm B-L}$ constraint that we consider here.
Importantly, adding a non-negligible number of neutrinos would only have the consequence of increasing the value of $\lambda$.
The baryon density fraction is then simply 
\begin{equation}
    \Omega_{\rm b} = \frac{\rho_{\rm baryons}}{\rho_c} \,. \label{eq:baryonfrac}
\end{equation}


\section{Confrontation with observations}\label{sec:observations}

We shall now compare our predictions described in Sec.~\ref{sec:predictions} with the cosmological measurements.
In particular we will constrain the model parameters required to match $\Omega_{\rm PBH}$ (Eq.~\ref{eq:BHdensity}) to the observed fractional energy density in CDM at early and late times as well as to reproduce the observed values for $\Omega_{\rm b}$ (Eq.~\ref{eq:baryonfrac}) and $Y_{\rm B-L}$ (Eq.~\ref{yield}) using various data sets. The free parameters we have available to predict these observables are the peak mass $M_{\rm pk}$ of the mass spectrum, the number density of PBHs today $\bar{n}_0$, the B-L current coupling $\lambda$, and finally the reheating temperature $T_{\rm RH}$. No sophisticated parameter inference analysis will be performed here as we simply wish to conduct a proof of concept and see if the predictions we derive may consistently capture the observations in principle. We begin with a brief discussion of the observational constraints.

\subsection{Observational constraints and parameter priors} \label{sec:priors}

For our anlysis we will consider three observational data sets, which we aim to match with our predictions. These are summarised in Table~\ref{tab:observations}.

\begin{table}
\centering
\caption{Observational constraints considered in our analysis.}
\begin{tabular}{| c | c | c | c | }
\hline  
 {\bf Parameter} & {\bf Data set } & {\bf Mean value}  &{\bf Assumption } \\
 \hline
 $H_{0}$ & \emph{Planck} 2018  & 67.32 & $\Lambda$CDM  \\ \hline 
$\Omega_{\rm CDM,0}$ & \emph{Planck} 2018  & 0.265 & $\Lambda$CDM  \\ \hline 
$\Omega_{\rm b,0}$ & \emph{Planck} 2018  & 0.049 &  $\Lambda$CDM   \\ \hline $\Omega_{\gamma,0}$ & \emph{Planck} 2018 & $9\times10^{-5}$ &$\Lambda$CDM \\ \hline 
$\Omega_{\rm CDM,0}$ & BOSS DR12 & 0.243 & \emph{Planck} and BBN priors  \\ \hline
$\Omega_{\rm b,0}$ & BOSS DR12 & 0.052 & \emph{Planck} and BBN priors  \\ \hline 
$Y_{\rm B-L}$ & BBN and SM &  $\geq 5.8 \times 10^{-10}$ & see App.~\ref{app:yieldobs}  \\ \hline 
\end{tabular}
\label{tab:observations}
\end{table}

The first data set are the CMB observations by the \emph{Planck} mission~\cite{Akrami:2018vks,Aghanim:2018eyx}, which provides a measurement of the composition of the Universe at early times.
From this we take the constraints inferred on the CDM and baryon density fractions, $\Omega_{\rm CDM}$ and $\Omega_{\rm b}$, from the imprints of the acoustic oscillations in the CMB.
The second will be the late-time measurement of the galaxy distribution from the BOSS survey~\cite{Ivanov:2019pdj}. These constraints on the matter fractions assume \emph{Planck} 2018 and BBN priors and so are not completely independent, but the results are consistent with other late-time data sets such as from the Dark Energy Survey (DES) weak gravitational lensing and photometric galaxy clustering~\citep{Abbott:2017wau}. 
Finally, we consider an observational bound on the baryon-lepton yield, $Y_{\rm B-L}$. This is slightly more complicated than the matter density fractions and so we have dedicated App.~\ref{app:yieldobs} to its discussion. We note here, however, that our observational constraint relies on both BBN physics and the Standard Model (SM) of particle physics. 

There are many mechanisms for PBH formation, for example, bubble collisions~\cite{Crawford:1982yz,Hawking:1982ga,Kodama:1982sf,Konoplich:1999qq} or collapse of scalar fields~\cite{Cotner:2018vug,Cotner:2019ykd}. It is usually assumed that the formation of PBHs happened from some time during inflation to the radiation-dominated epoch~\cite{Leach:2000yw,Ozsoy:2020kat}. In accordance with Eq.~\eqref{eq:mf}, we can estimate that the PBHs formed during inflation have masses of roughly $M_p$ whereas those formed at the end of radiation domination are roughly of order $10^{46}$~kg.
However, many studies~\cite{Barrau:2003xp,Montero-Camacho:2019jte} suggest that the mass of PBHs as dark matter candidates would be bounded by a range of observational constraints, leaving only a few narrow mass windows: the asteroid-mass range ($10^{13}$--$10^{14}$~kg), the sublunar-mass range ($10^{17}$--$10^{23}$~kg), and the intermediate-mass range ($10^{30}$--$10^{33}$~kg).
To explore our model, we adopt a broad mass range of $M_p \leq M \leq 10^{36}$~kg, which is the prior we place on the peak mass $M_{\rm pk}$ and also the range over which we perform the integrals in Eqs.~\eqref{eq:avgmass} and \eqref{totalnbl}.

Besides the prior on $M_{\rm pk}$, we also consider priors on the other free parameters of our framework.
We have no strong constraints on the number density of PBHs, except that we  want it to be much larger than the number density of black holes formed from stellar collapse, which could not account for all CDM and in which case also the approximation of a fixed number of black holes we made would break down (see App.~\ref{app:assext}).
Furthermore, at the time of recombination we do not want the number density to be anywhere close to the maximal packing limit, represented by a volume fraction of $\approx 0.74$, a result of Carl F.~Gauss in 1831.
This should not be of concern, however, as otherwise any dark matter model would fail.
Nevertheless, we shall briefly inspect this constraint,
\begin{equation}
   \left(\frac{ V_{\rm PBH}} {V_{\rm tot}}\right)_{\rm CMB} = \bar{n}_{\rm CMB} \frac{4}{3} \pi r_+^3 = \bar{n}_{\rm CMB} \frac{32}{3} \pi  \langle M(a_{\rm CMB}) \rangle^3 \ll 0.74 \,,
\end{equation}
where CMB denotes the time of last scattering.
This translates to $\bar{n}_{\rm CMB} \ll 5 \times 10^{43}~{\rm m}^{-3}$ assuming Schwarzschild black holes with masses equal to the average mass $\langle M (t_{\rm CMB}) \rangle$ and a monochromatic mass spectrum with peak mass of $M_{\rm pk} = 10^{12}$~kg. 

Finally, the reheating temperature is constrained to $10^{-4}{\rm GeV} \leq T_{\rm RH} \leq 10^{12} {\rm GeV}$, considering that reheating needs to occur after the end of inflation, defining the maximum temperature, and before the BBN, defining the minimum temperature (see Eq.~\ref{eq:trh})~\cite{Hannestad:2004px,Kawasaki:2000en}.

Note that we take no prior on the coupling $\lambda$. In principle this can take any value given our ignorance of a full quantum gravity theory. We refer the reader to Ref.~\cite{Hook:2014mla} for a discussion of why it can particularly assume arbitrarily large values.


\subsection{Results}

\begin{figure}[h]
\centering
  \includegraphics[width=0.85\textwidth]{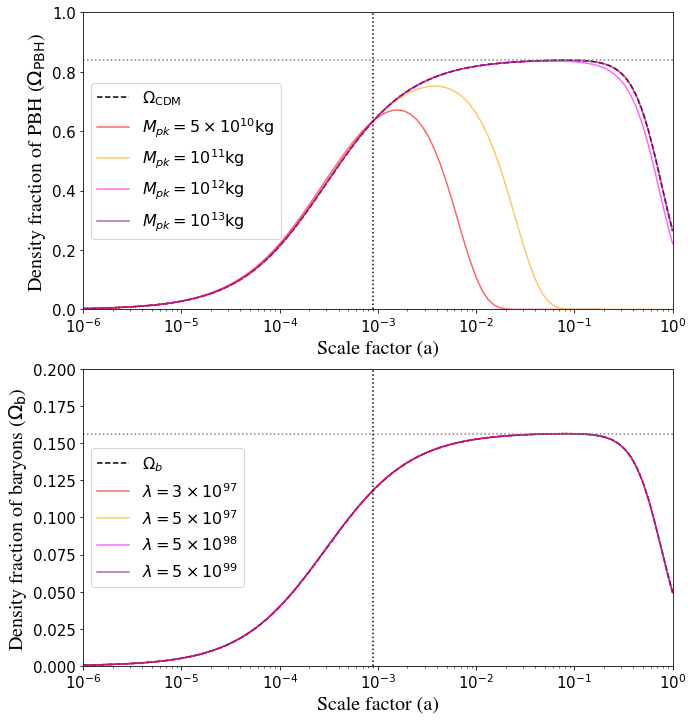}
  \caption[]{Fractional energy densities for PBHs (top panel, Eq.~\ref{eq:BHdensity}) and for baryons (bottom panel, Eq.~\ref{eq:baryonfrac}) for $\psi_{\rm mono}$ with various choices of peak masses. We tune $\bar{n}_0$ and $\lambda$ to match the \emph{Planck} 2018 $\Lambda$CDM mean values for $\Omega_{\rm CDM}$ and $\Omega_b$ at the time of recombination, $a_{\rm CMB} = 9\times 10^{-4}$, marked as a vertical dotted line. The $\Lambda$CDM predictions for $\Omega_{\rm CDM}$ and $\Omega_{\rm b}$ are given by dashed black lines. The horizontal gray dotted lines indicate the observed proportions of total matter in the CDM and baryon fractions. The reheating temperature is taken to be $T_{\rm RH} = 10^{8}$~GeV. } 
\label{fig:densityfrac}
\end{figure}

We can now turn to investigate the capacity of our setup to reproduce the desired $\Lambda$CDM density fractions at the last-scattering surface and today.
To do this, we begin by selecting a peak mass for the mass spectrum (see Eq.~\ref{eq:mono}). Once chosen, we then tune the number density of PBHs in Eq.~\eqref{eq:BHdensity} to match the \emph{Planck} 2018 measurement of $\Omega_{\rm CDM}$ at the time of recombination.
In the top panel of Fig.~\ref{fig:densityfrac} we show $\Omega_{\rm PBH}$ as a function of the scale factor for the monochromatic spectrum.  We find that although we can always match $\Omega_{\rm CDM}$ at the time of recombination, $\Omega_{\rm PBH}$ will vanish by today unless $M_{\rm peak}\geq 10^{12}$~kg.
This is due to the evaporation of the PBHs and the consequent vanishing average mass.
In contrast, for the polychromatic case, we can recover the correct evolution for all choices of peak masses. Here, the evolution of the average mass with time is negligible after some very early epoch, and the correct $a^{-3}$ evolution is produced post recombination. 

\begin{figure}[h]
\centering
  \includegraphics[width=0.85\textwidth]{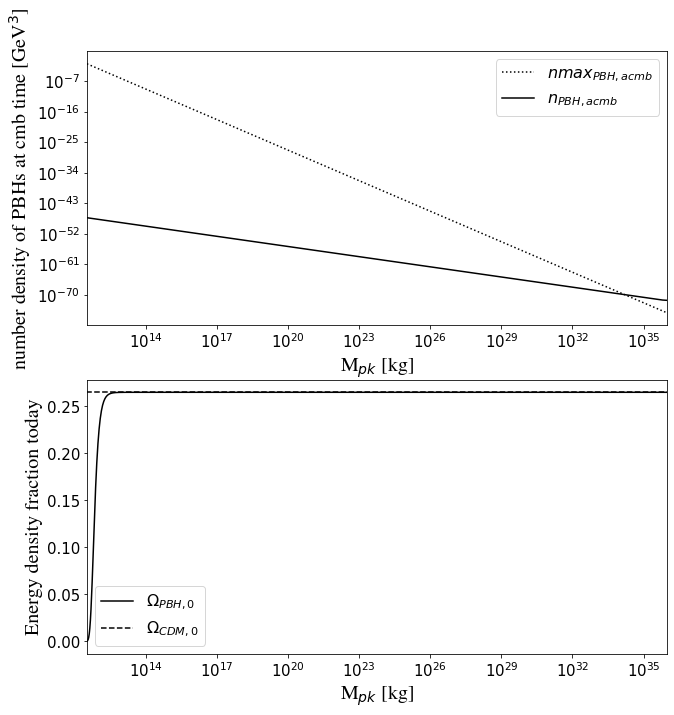}
  \caption[]{The top panel shows the number density of PBHs at the time of recombination $n_{\rm PBH, acmb}$ as a function of peak mass $M_{\rm pk}$. The dotted line represents the geometrically maximally allowed number density for the mass spectrum at the time of recombination. The bottom panel shows the fractional energy density predicted for PBHs with $\psi_{\rm mono}$ at present day, $\Omega_{\rm PBH,0}$, using Eq.~\eqref{eq:BHdensity}, as a function of peak mass $M_{\rm pk}$. We tune $\bar{n}_{\rm PBH, 0}$ to match the $\Lambda$CDM prediction for $\Omega_{\rm CDM}$ at the time of recombination as in Fig.~\ref{fig:densityfrac}. The dashed line represents the fractional energy density of CDM at present day according to \emph{Planck} 2018.}
\label{fig:npbh_omegapbh_peakm}
\end{figure}

Furthermore, we find that for a peak mass of $10^{12}$~kg, the monochromatic spectrum requires a current number density of $\bar{n}_0 \sim  10^{-39}~{\rm m}^{-3}$ whereas the polychromatic spectrum, which has a naturally larger average mass, requires  $\bar{n}_0 \sim \times 10^{-51}~{\rm m}^{-3}$. We show the number density required to match $\Omega_{\rm CDM,CMB}$ as a function of peak mass for the monochromatic spectrum in the top panel of Fig.~\ref{fig:npbh_omegapbh_peakm}. We see for $M_{\rm pk} \leq 10^{34}$~kg that we are well below the prior bounds on the number density at time of recombination discussed in Sec.~\ref{sec:priors}. The density fraction today, $\Omega_{\rm PBH,0}$ as a function of peak mass, fixing $\bar{n}_0$ to match $\Omega_{\rm CDM}(a_{\rm CMB})$, is shown in the bottom panel of Fig.~\ref{fig:npbh_omegapbh_peakm}. This also shows that for all $M_{\rm pk} > 10^{12}$~kg we can recover the correct CDM density today. 
Note that for the approximation we use in Eq.~\eqref{eq:lcdmback} to be valid, we require peak masses to be around this bound or larger so that the expansion history is correctly represented by $\Lambda$CDM, but also so that the entropy density $s$ and the PBH number density $n_{\rm PBH}$ have the same time evolution, needed in Eq.~\ref{eq:blasym}.

Once we have fixed $M_{\rm pk}$ and $\bar{n}_0$, the next parameter to inspect is the reheating temperature $T_{\rm RH}$. This parameter dictates when we begin integrating the charge in Eq.~\eqref{eq:qblint} for the monochromatic case. A higher reheating temperature implies an earlier time and so we have more time to generate the required amount of baryonic matter before recombination. Here we set $T_{\rm RH} = 10^8$~GeV which is well above the BBN temperature of 4~MeV and below the temperature at the end of inflation of $\sim 10^{\rm 12}$~GeV.
However, irrespective of what value this takes, there is a direct degeneracy with $\lambda$ (see Eqs.~\ref{eq:qtimdep} and \ref{eq:chempot} where $a_{\rm i}=a_{\rm RH}$). Once we have fixed the reheating temperature, the value of $\lambda$ is uniquely determined if we wish to match the mean value of \emph{Planck} 2018 for $\Omega_{\rm b, CMB}$. We show the evolution of $\Omega_{\rm b}$ for various choices of peak masses in the bottom panel of Fig.~\ref{fig:densityfrac} for the monochromatic case.

\begin{figure}[h]
\centering
  \includegraphics[width=0.85\textwidth]{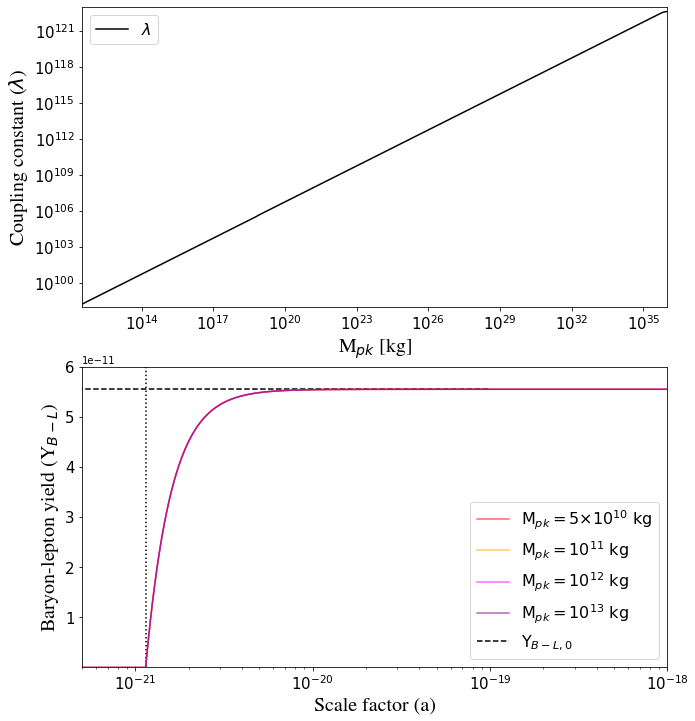}
  \caption[]{The top panel shows the coupling constant $\lambda$ required to match predictions for $\Omega_b$ to the observed fractional energy density of baryonic matter at the time of recombination as a function of the peak mass $M_{\rm pk}$ of $\psi_{\rm mono}$. The bottom panel shows the baryon-lepton yield $Y_{\rm B-L}$ as a function of the scale factor $a$ for a reheating temperature in accordance with Eq.~\eqref{yield}. The dotted vertical line corresponds to the scale factor at reheating time $a_{\rm RH}$ and the dashed horizontal line corresponds to the predicted yield at present day. The reheating temperature is taken to be $T_{\rm RH} = 10^{8}$~GeV.}
\label{fig:lambda_ybl_peakmanda}
\end{figure}

Let us make a few remarks here.
Firstly, note that $\lambda$ must assume exceedingly high values to match the observed baryon density. It increases with $M_{\rm pk}$ since for larger peak masses we have a lower number of PBHs and hence less total evaporation. We show its dependence with peak mass in the upper panel of Fig.~\ref{fig:lambda_ybl_peakmanda} for the monochromatic spectrum. It also naturally decreases with increasing $T_{\rm RH}$ since a larger reheating temperature gives more time to produce baryons and so the coupling does not need to be as strong. Finally, we find that the evolution of $\Omega_{\rm b}$ goes almost exactly as $a^{-3}$ independently of peak mass or reheating temperature. This can be seen from Eq.~\eqref{totalnbl}. Apart from the scaling with $a^{-3}$ we have a time dependence in the charge $Q(M_i,a_*)$ given by Eq.~\eqref{eq:qtimdep}. However, this is just a constant for times sufficiently past the end of reheating.

In contrast, for the polychromatic case, we have no dependence on the reheating temperature in $\Omega_b$ since the PBHs are produced in the radiation-dominated epoch. In this case, we require much higher values of $\lambda$ to match the observations despite having a larger average mass for the same peak mass. This is because we begin producing the asymmetry at later times than in the monochromatic case. For reference, for a peak mass of $M_{\rm peak}=10^{12}$~kg we need $\lambda \sim 10^{99}$ and $\lambda\sim 10^{134}$ for the monochromatic and polychromatic cases respectively. 

Having fixed all free parameters we are left with a unique determination of the yield. This is found to be $Y_{\rm B-L} = 5.6 \times 10^{-11}$ at the time of BBN and for later times, both for the monochromatic and polychromatic spectra. This is an order of magnitude less than the lower bound we derive in App.~\ref{app:yieldobs}, $Y_{\rm B-L} \geq 5.8  \times 10^{-10}$. We show the evolution of the yield $Y_{\rm B-L}$ in the bottom panel of Fig.~\ref{fig:lambda_ybl_peakmanda} for the monochromatic case. The yield is dependent on $n_{\rm B-L}$ and the entropy density $s$. The $a^{-3}$ dependence of each of these quantities cancel and we get a quantity that goes as  $(T_{\rm RH} \times a_{\rm RH})^{-3}$ and has an evolution specified by Eq.~\eqref{eq:qtimdep}, making it constant for times following reheating.

Let us make a few further remarks here.
To change the prediction for $Y_{\rm B-L}$ we must change the entropy density since  $n_{B-L}$ is fixed to obtain the correct $\Omega_b$ given by CMB observations. To do this, the background evolution becomes important. 
Changing the background at early times can be achieved by considering, for example, an early matter-dominated phase or a scenario where reheating is caused by the PBHs. On the other hand, if we consider loop corrections to the running of gauge coupling constants, changing $w$ in radiation domination, will give us a non-vanishing chemical potential even during radiation domination~\cite{Kajantie:2002wa}, but will not significantly help increase the yield.

\begin{figure}[h]
\centering
  \includegraphics[width=0.85\textwidth]{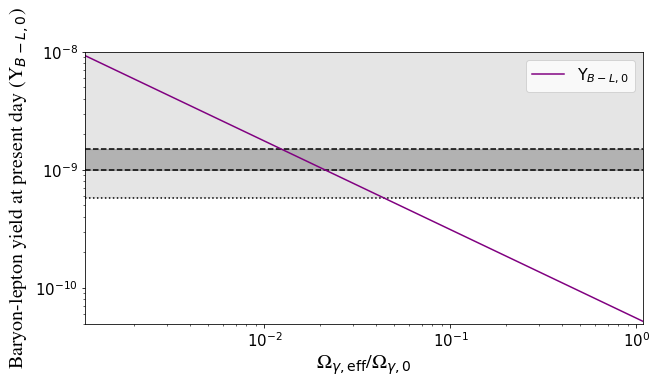}
  \caption[]{Baryon-lepton yield $Y_{\rm B-L}$ as a function of the ratio between the effective fractional radiation energy density extrapolated from early to present times $\Omega_{\gamma,eff}$ and the observed fractional radiation energy density today $\Omega_{\gamma,0}$.
  If a significant amount of the radiation is produced by the decay of PBHs, $\Omega_{\gamma,eff} \ll \Omega_{\gamma,0}$, as would for instance be expected if the decaying PBHs were to cause reheating, the model can reproduce the observed yield. The dotted and dashed lines are observational bounds for different assumptions (App.~\ref{app:yieldobs}). The reheating temperature is taken to be $T_{\rm RH} = 10^{8}$~GeV.} 
\label{fig:yield_rad}
\end{figure}

Expanding on this, the quantity $T_{\rm RH} \times a_{\rm RH}$ in the entropy density $s$ (see Eq.~\ref{eq:entropydens}) turns out to be a constant since at early enough times we are radiation dominated and so $t \propto a^2$. Then by Eq.~\eqref{eq:trh} we get $T_{\rm RH} \propto a_{\rm RH}^{-1}$. This makes the yield  sensitive to the radiation density. A different density fraction of radiation at early times from that na\"ively extrapolated from the recombination backwards with $a^{-4}$ in our approximation will change the scale factor of reheating $a_{\rm RH}$ in Eq.~\eqref{eq:entropydens} and hence the yield.
Note that it is clear that the $a^{-4}$ dependence ceases to hold when non-relativistic particles become relativistic at high temperatures.
To investigate this dependency we vary the fractional energy density of radiation $\Omega_{\gamma,0}$ in Eq.~\eqref{eq:lcdmback} as an effective parameter $\Omega_{\gamma, eff}$.
This variation will mainly affect the total energy density in the radiation epoch.
Tuning $\Omega_{\gamma, eff}$ then implies an effective correction to the $a^{-4}$ extrapolation.
We only wish to see the effects on the entropy density shortly after reheating (see bottom panel of Fig.~\ref{fig:lambda_ybl_peakmanda}).
We show this dependency in Fig.~\ref{fig:yield_rad}.
For an excess radiation density, and so more radiation in the early Universe than na\"ively extrapolated from recombination
we obtain a decrease of the predicted yield.
If on the other hand we have less radiation in the early Universe than extrapolated from recombination we find an increase in the predicted yield.
This effect could potentially reconcile the model with the observed yield.
In particular, we would require $\sim 95\%$ of the observed radiation density to be produced at some point during or after reheating. One possible scenario is that the decaying PBHs themselves produce this radiation, which is also what one would expect if PBHs were to be responsible for reheating~\cite{Hook:2014mla}.

More generally, a $\Lambda$CDM expansion will not be valid at very early times, where matter couplings and even the inflaton become relevant. We refer the reader to App.~\ref{app:assext} for a discussion on the various assumptions we have made and their impact on the results.
An in-depth analysis of these effects on the prediction of the yield is, however, beyond the scope of this initial proof-of-concept paper.

\section{Conclusions}\label{sec:summary}

We have investigated the potential of primordial black holes (PBHs) of distributed masses to match the observed cold dark matter (CDM) abundance at early and late times, while also explaining the observed baryon-antibaryon asymmetry and baryon abundance through an asymmetric Hawking radiation mechanism.

We find that our predictions for the PBH and baryon density fractions can be made consistent with the CDM and baryon fractions inferred from \emph{Planck} 2018 measurements of the CMB and from late-time measurements of the large-scale structure with BOSS.
Furthermore, we find that a simplified prediction for the amount of matter-antimatter asymmetry, characterised by the effective baryon-lepton yield, can also be made consistent with observations coming from Big Bang nucleosynthesis to within one order of magnitude. These results hold for both narrow (monochromatic) and broad (polychromatic) mass distributions of PBH.
We argued furthermore that the exact observed value of the baryon-lepton yield can in principle be recovered in a scenario where a significant amount of the radiation at early times is produced by the PBH decay during or after reheating, which would be what is expected if the decaying PBHs were also to be responsible for reheating.

Matching the $\Lambda$CDM evolution requires large peak masses ($M_{\rm pk}\geq10^{12}$~kg) for the monochromatic mass spectrum, which then implies very large coupling constants to generate the required baryon density fractions.
This is because massive PBHs generate less Hawking radiation.
The mechanism producing the non-vanishing baryon number is subject to quantum gravity corrections and so it is difficult to make any comments on the physicality of the magnitude of the coupling. For the polychromatic spectrum, we can recover a $\Lambda$CDM evolution for all peak masses since there is always a population of PBHs that has not evaporated. By tuning the coupling constant and number density of PBHs, we can recover the correct observed values for the fractional matter densities.

In future work, we aim to extend our framework to include a more accurate modelling of the background expansion deeper into the early Universe.
This will provide a more accurate computation of the yield and also allow to explore the effects of an early matter-dominated phase following inflation on this computation. Furthermore, we will also examine the ability of PBHs to fuel reheating which was already considered in Ref.~\cite{Hook:2014mla}, albeit for a single PBH mass population. 

All calculations included in this work were made using the {\tt c++} code made publicly available at \url{https://github.com/nebblu/PBH}.

\section*{Acknowledgements}

We would like to thank Valerio de Luca and Gabriele Franciolini for useful discussions and valuable feedback on the manuscript. A.B.~acknowledges the support by an international mobility scholarship from the University Claude-Bernard (ARI) and the University of Geneva (Movetia). B.B.~and L.L.~acknowledge the support by a Swiss National Science Foundation
(SNSF) Professorship grant (No.~170547). H.H.~gratefully acknowledges the support by an international scholarship from Beijing Normal University for a research visit to the University of Geneva.

\appendix

\section{Further details} \label{sec:details}

For completeness we shall now provide further details of our computations in Secs.~\ref{sec:model}--\ref{sec:observations}, specifically the effective degrees freedom used (Sec.~\ref{sec:dof}) and the observational constraints considered on the yield (Sec.~\ref{app:yieldobs}).

\subsection{Effective degrees of freedom}\label{sec:dof}

We have made use of the total number of relativistic degrees of freedom in our Universe in our computations of Secs.~\ref{sec:model}--\ref{sec:observations} such as for the decay of black holes, the baryon-lepton charge, or the entropy density.
We provide a brief inspection of this quantity here. 
We shall first take into account the particle degeneracies only, in which case one finds
\begin{eqnarray}
        \sum_i g_i q_i^2 & = & \sum_{B-L \: charged \: particles} g_i q_i^2 \, \nonumber\\
         & = & \Big(\sum_{quarks} + \sum_{charged \: leptons} + \sum_{neutrinos}\Big) g_i q_i^2 \, \nonumber\\
         & = & 36 \times \Big(\frac{1}{3}\Big)^2 + 6 \times 1^2 + 3 \times 1^2 \, \nonumber\\
         & = & 4 + 6 + 3 \, \nonumber\\
         & = & 13 \,.
\end{eqnarray}
Taking into account both particle and anti-particle degeneracies (see Table~\ref{tab:degeneracy}), one finds
\begin{eqnarray}
        g_\star & = & \frac{7}{8}\sum_{fermions}g_i + \sum_{bosons}g_i \, \nonumber\\
        & = & \frac{7}{8}(72 + 12 + 6) +(16 + 2 + 9 + 1)\, \nonumber\\
        & = & 106.75\, .
\end{eqnarray}  
These two quantities change over time, but only the second one, the effective SM number, has an impact on our computation of the yield $Y_{B-L}$ because of the dependence of the entropy on it.
The number varies significantly between the early Universe and the present.
To effectively model the impact of this change, we have introduced the parameter $\Omega_{\gamma,eff}$ in Sec.~\ref{sec:observations}.
However, we have left a more accurate computation of this effect to future work as it is complexly intertwined with the amount of relativistic radiation emitted by the decay of the PBHs, which requires a more sophisticated computation and lies beyond the scope of our current work.

\begin{table}[h]
\centering
\caption{Summary of Standard Model particle degeneracies with antiparticles at high temperatures $T\geq 100$~GeV at early times. For more details, we refer to Table~1 of Ref.~\cite{Husdal_2016}.}
\begin{tabular}{| c | c |}
\hline  
 {\bf Particles} & {\bf Total degeneracies}  \\
 \hline
 Quarks & 72 \\ \hline 
 Charged leptons  &  12  \\ \hline 
 Neutrinos &  6 \\ \hline 
 Gluons & 16 \\ \hline
 Photons &  2 \\ \hline 
 Massive gauge bosons & 9 \\ \hline 
 Higgs bosons & 1 \\ \hline 
All elementary particles & 118 \\ \hline 
\end{tabular}
\label{tab:degeneracy}
\end{table}

\subsection{Observational constraints on the yield} \label{app:yieldobs}

In Fig.~\ref{fig:yield_rad} we have compared our predicted yield against cosmological constraints.
In the following we shall briefly specify the origin of these constraints.
Firstly, BBN bounds and CMB observations imply that the baryon number asymmetry should lie in the range of $5.8 \times 10^{-10} \leq B \leq 6.5 \times 10^{-10}$ (95\% C.L.)~\cite{Cyburt_2016, Planck2016, PhysRevD.98.030001}.
The SM assumes a lepton number asymmetry of the same order as the baryon number asymmetry, and we will limit ourselves to this case here, thus $L \approx -B$. The baryon-lepton yield is then constrained to lie in the range of
\begin{equation}
    1.0 \times 10^{-9} \leq Y_{\rm B-L} = B - L \leq 1.5 \times 10^{-9}\, , \label{eq:const1}
\end{equation}
admitting a possible small difference between the value of $L$ and the value of $B$.

However it is important to keep in mind that the SM has gaps and that it is not impossible, especially with future advances in neutrino physics, that other scenarios which produce a different asymmetry will become predominant~\cite{Barenboim_2017_2, Caramete_2014, yang2018lepton}. Depending on the model, $Y_L$ will either be smaller than $Y_B$, even negligible some times, or higher than $Y_B$. The only model independent bound is
\begin{equation}
    Y_{\rm B-L} \geq 5.8 \times 10^{-10} \,.
    \label{eq:broaderconstraintonY}
\end{equation}

In Fig.~\ref{fig:yield_rad}, we have considered Eq.~\eqref{eq:const1} as our main constraint on $Y_{\rm B-L}$ and Eq.~\eqref{eq:broaderconstraintonY} as a secondary bound.
Interesting scenarios therefore are ones that respect at least one of the two constraints.

A further restriction one may impose is that the majority of the asymmetry allocated to the baryons must be present before the BBN process occurs if we want our predictions to be consistent with the abundances observed today. The yield should therefore not change significantly between BBN and recombination and even less between recombination and the present day~\cite{sarkar2002measuring, PhysRevD.98.030001}. But it is important to note that this constraint supposes that all the baryonic matter is already present at BBN, which is not necessarily the case for the models studied here.
Taking into account the possible changes in BBN is a complex task and for a first proof-of-concept study we only took into account CMB constraints, deferring BBN bounds to a later analysis.


\section{Assumptions and Extensions}\label{app:assext}

Finally, in the following, we shall provide a summary of the various assumptions we have made and their expected impact on the final calculations presented in Sec.~\ref{sec:predictions}.
We will also discuss extensions of our framework to be considered for future work.

\paragraph{Stellar collapse, mergers, and accretion} 

We have assumed that the total number of black holes formed remains constant since the early Universe and that new black holes formed at late times do not provide a significant contribution.
We expect this to be a very good approximation as the abundance of PBHs considered is such that it constitutes all of the dark matter.
The relative contribution of conventionally formed black holes at late times is therefore of the same magnitude as for usual dark matter models and can thus safely be neglected.
More specifically, taking a constant number of PBHs allows us to fix the distribution of initial masses given by Eq.~\eqref{eq:massspecdef}. This is in principle not exact as black holes will have formed due to stellar collapse, mergers, and matter accretion producing a larger number of black holes and extending the mass range considered here ($M_p \leq M \leq 10^{36} $~kg). Note that stellar-mass black holes are unable to account for a significant amount of CDM due to the low number density and mass ranges and so can be neglected. 
The effect of accretion will work to boost the average mass at late times by broadening the mass spectrum~\cite{DeLuca:2020fpg}. This may allow us to assume smaller peak masses (depending on the specifics of the accretion), and so act to boost baryogenesis at early times. We aim to consider this in a future work. Mergers on the other hand will sharpen the spectrum and move the peak towards higher masses. The number of mergers is expected to be negligible due to the low number density at recombination\footnote{For $M_{\rm pk} = 10^{13}$~kg we need $\bar{n}_{\rm CMB} \approx 10^{-42}~{\rm m}^{-3}$ to match the required CDM density from \emph{Planck} 2018.}.

\paragraph{Effect of PBH remnants}

We have assumed that black holes completely decay through Hawking radiation, leaving no stable remnants behind. We have checked that the impact of stopping decay once the PBHs reach the Planck mass and find that this only slightly affects the values of $\bar{n}_0$ for both the monochromatic and polychromatic mass spectrum.

\paragraph{Background expansion}

In Ref.~\cite{Hook:2014mla} the authors take the evolution of the Universe as being divided into fully separated epochs, in which a single energy density component dominates, and so $w\approx {constant}$.
This is a rough approximation, which we only partially improve by our adoption of the $\Lambda$CDM background expansion. A more accurate treatment would require solving the set of coupled differential equations 
\begin{eqnarray}
       Q_{\rm B-L}' = & -\bar{\lambda}H^2[(1+w)(1-3w)+w'] \,, \\
       H' = & H_0 \left( \sqrt{\Omega_{\rm PBH}(a) + \Omega_{b}(a) + \frac{\Omega_{\gamma,0}}{a^4} + \Omega_{\phi}} \right)' \,,
\end{eqnarray}
where $\Omega_\phi$ is the inflaton fractional energy density and the dark energy contribution to $H$ can be ignored at early times. We have defined here $\bar{\lambda} \equiv 9\lambda \sum_i q_i^2 g_i /(96 \pi M_p^2)$. The PBH energy density evolution is specified by Eq.~\eqref{eq:BHdensity} while the baryon energy density is given by Eq.~\eqref{eq:bardens}. 
The first set of initial conditions can be taken as  $H_i = H_{\rm inf} =10^{14}$~GeV \cite{Ade:2014xna}, $Q_{\rm B-L,i}=0$ (due to inflaton domination). The initial derivatives can be derived assuming that after inflation we are dominated by the inflaton ($\phi$), PBHs, and radiation. This then depends on the free parameter $\bar{n}_{0}$ through Eq.~\eqref{eq:BHdensity}. We write the explicit set of initial conditions as  $\{ H_i, Q_{\rm B-L,i} \} =  \{10^{14}, 0 \}$ and 
\begin{eqnarray} 
       Q'_{\rm B-L,i} = &  -\bar{\lambda}H_i^2[(1+w_i)(1-3w_i)+w'|_{H_i}] \,, \\ 
       H'_{i} = & H_0 \left( \sqrt{\Omega_{\rm PBH}(\bar{n}_0) + \Omega_{\phi}  + \Omega_\gamma}\right)' \,.
\end{eqnarray}
The initial time is just given by $t_i = \frac{2}{3}H_i$. Note that we do not expect this approach to change our results significantly since we require large peak masses to match observations, and so the evolution of PBHs is very close to that of CDM.
Importantly, this computation still neglects the change in the radiation density at early times when matter species become relativistic.
Increasing the radiation density will increase the scale factor at which reheating ends for the same temperature (see Eq.~\ref{eq:trh}) and at which we begin producing baryons. Thus, more radiation reduces the yield whereas less radiation will increase it. On the other hand, an increase in the Hubble constant, $H_0$, will act to increase the yield, albeit marginally within the uncertainty given by \emph{Planck} 2018.

\paragraph{Chemical potential}

We note that the form of the chemical potential in Eq.~\eqref{eq:chempot} and its evolution will also affect when and how much asymmetry and baryons are produced as well as their time dependence (see Eq.~\ref{eq:qtimdep}). For example, in Ref.~\cite{Hamada_2017}, the authors consider a chemical potential generated from the PBH evolution rather than from the expansion of the Universe, which has a significantly different form and time dependence.
An interesting generalisation of our work would be to use a parametrisation of the chemical potential that can represent a broader range of generation processes.

\paragraph{Initial matter domination phase}

Another important consideration is the presence of a pre-BBN matter dominated epoch, a situation common to low-energy limits of supergravity or M-theory compactifications (see for example Ref.~\cite{Easther:2013nga}). Such an era would trigger asymmetric radiation (see Eq.~\ref{eq:chempot} with $w=0$) and boost the value for the B-L yield we obtain in Sec.~\ref{sec:predictions}. This would involve a transition to a more complicated background expansion at early times with couplings between massive scalar particles, SM particles, and the inflaton~\cite{Drees:2017iod}. These complications are beyond the scope of this preliminary study and hence we leave such an analysis to future work. 
Considering this additional matter domination phase allows to add a dependence on the reheating temperature in the entropy definition. Indeed, $s^{1/3} \propto T_{RH} \times a_{RH}$ at a given time and using a $\Lambda$CDM background we obtain that $s^{1/3} \propto constant$ as discussed in Sec.~\ref{sec:predictions}. With the matter domination phase before reheating we can expect the reheating to occur during the corresponding matter-radiation equality which will change the Hubble parameter at reheating time and allow $s^{1/3} \neq constant$.

\paragraph{Reheating from PBHs}

One can also consider the decay of PBHs as the source of reheating~\cite{Hook:2014mla}. In this case, one would want to have
\begin{equation}
 \bar{n}_i \langle M(t_{\rm RH}) \rangle = \frac{\pi^2}{30} g_\star T_{\rm RH}^4 \,,
\end{equation}
where $t_{\rm RH} = \sqrt{\frac{5}{\pi^3g_\star}}\frac{M_p}{T_{\rm RH}^2}$ and $M(t)$ represents the mass lost by a PBH since $t_i$. We can solve this for $T_{\rm RH}$, or $t_{\rm RH}$, and make sure that reheating happens well before recombination. We aim to investigate this in a future work.
As discussed in Sec.~\ref{sec:observations} such a process may furthermore increase the baryon-lepton yield to an observationally compatible level.

\paragraph{Dependence on the maximum mass}

Finally, in this work we have fixed a maximum mass of $M_{\rm max} = 10^{36}$~kg, or correspondingly the longest wave mode to re-enter the horizon, in calculating various quantities in Secs.~\ref{sec:model}--\ref{sec:observations} (e.g. Eq.~\ref{eq:avgmass} for the average PBH mass), which assumes that PBHs can be created up to the end of reheating. Reheating can at most last until just before BBN~\cite{Coc:2017pxv}. This is particularly relevant for the polychromatic spectrum \cite{DeLuca:2020ioi} for which we can continue to produce black holes into the radiation dominated epoch ($t>10$~s).  For the monochromatic spectrum we assume that the spectrum is fixed at the end of inflation, which is inconsistent with the further formation of PBHs. 
In effect, this results in a different maximum mass when integrating Eq.~\ref{eq:avgmass}. Despite this, our results are currently insensitive to this choice of $M_{\rm max}$ as long as it is not close to the peak mass which is the case in this work. The choice of $M_{\rm max}$ is also perfectly degenerate with $\bar{n}_0$ for $\Omega_{\rm PBH}$ for which we have no lower bound, and similaly with $\lambda$ for $\Omega_b$, upon which we place no priors. In future work, the choice of reheating temperature should be related to this maximum mass since it roughly governs the latest time at which PBHs can form.

\bibliography{mybib} 
\bibliographystyle{ieeetr}

\end{document}